# Hydrogen ordering influence on the proton spin-lattice relaxation time in metal-hydrogen compounds

G.Mamniashvili[a], N.Namoradze[b] I.Ratishvili[a], Yu.Sharimanov[a*]

[a] E.Andronikashvili Institute of Physics, Georgian Academy of Sciences
[b] E.Institute of Cybernetics, Georgian Academy of Sciences

*Corresponding author
E-mail address: usharimanov@iphac.ge
(Yu.G.Sharimanov).

## Abstract

Results of the experimental study of ordering lanthanum hydrides of $LaH_X$ system are presented. It is shown that obtained results can be interpreted as an indirect confirmation of the excistence of order-order transformations in $LaH_{2+c}$ hydrides.

**Introduction**

*Brief review of the problem under consideration.* It is known that in metal-hydrogen interstitial solid solutions and in the hydride phases the proton spin-lattice relaxation time $T_1$ is determined by several mechanisms, among which the main roles play interactions with conduction electrons, with paramagnetic impurities and that caused by the dipole-dipole contacts of H-atoms during their diffusion within the lattice of surrounding nuclear spins [1-3]. It is known as well that the dominant role of the nuclear dipole-dipole interaction is revealed in the temperature range where the $T_1(T)$ dependence exhibits a pronounced minimum. Outside of this region, at high temperatures, as well as at low temperatures one of the frequently mentioned channels of the nuclear spin-system energy dissipation is the contact with conduction electrons (see e.g. [4, 5]), but as it was shown in [3] an excellent description of the experimental $T_1(T)$ dependence can be obtained by taking into account interactions of interstitial protons with impurity ion spins.

It has to be noted that diffusion of H-atoms in the metal lattice is characterized usually by a single activation energy $E_a$ defining the hoping rates of individual H-particles, which provides a simple $T_1(T)$ dependence with a single minimum. The corresponding experimental curves are more complicated and on trying to describe the real $T_1(T)$ dependences investigators made a number of attempts to subdivide the mobile protons into the separate groups characterized by different activation energies [2, 5, 6].

From our point of view, the ordering processes developed in the hydrogen subsystem of several metal hydrides can be considered to be responsible for a subdivision of interstitial hydrogen atoms into the groups characterized by different activation energies. To verify this idea we choused such a well known hydride as $LaH_x$ (see e.g. [7], [8]) and reinvestigated it in the temperature range where the ordering processes are known to be developed [9].

*The system under consideration.* In compounds $LaH_{2+c}$ ($0 < c < 1$) N metal atoms form a fcc lattice. The existing 2N tetrahedral interstitial sites are completely filled by 2N hydrogens denoted below as $H_T$-atoms. The remained $cN$ hydrogen atoms (called $H_O$-atoms) are distributed among N octahedral interstitial sites. At low temperatures the subsystem of $H_O$-atoms undergoes disorder-order and order-order phase transitions and forms different ordered configurations described by one or two long-range-order parameters, $\eta_1$ and $\eta_2$ [9]. The $H_O$-atoms number increase causes the metal lattice contraction, while the developed ordering processes provide the cubic metal lattice distortions [10].

*Description of the spin-lattice relaxation due to the dipole-dipole interactions.* The spin-lattice relaxation caused by dipole-dipole interactions of the given H-atom with surrounding protons (nuclear spin $I = 1/2$) and lanthanum nuclei (nuclear spin $S = 7/2$) is characterized by the corresponding relaxation time $T_{1d}$ and can be determined basing on the BPP scheme. In the case of variable field with frequency $f_0$ it follows [2]:

$$T_{1d}^{-1} = (\gamma_H^2/\omega) [(2/3) f_1(y) M_{HH} + f_2(y) M_{HM}], \qquad (1)$$

where $\omega = 2\pi f_0$, $y = \omega\tau$,

$$\tau = \tau_0 \exp(E_a / kT), \qquad (2)$$

$$f_1(y) = y [(1/(1+y^2)) + 4 (1 / (1 + 4 y^2))], \qquad (3)$$

$$f_2(y) = y [(1/(1+y^2)) + (1/3)(1/(1 + k_1 y^2)) + 2 (1/(1+k_2 y^2))], \qquad (4)$$

$$k_1 = (1 - (\gamma_M/\gamma_H))^2, \qquad k_2 = (1 + (\gamma_M/\gamma_H))^2, \qquad (5)$$

$$M_{HH} = (3/5)(h/2\pi)^2 \gamma_H^2 I(I+1) [\Sigma_j^{H(o)} n_j (R_{0j})^{-6} + \Sigma_j^{H(T)} (R_{0j})^{-6}], \qquad (6)$$

$$M_{HM} = (4/15)(h/2\pi)^2 \gamma_M^2 S(S+1) \Sigma_j^M (R_{0j})^{-6} \qquad (7)$$

In the above expressions $\gamma_H$ and $\gamma_M$ are the gyromagnetic ratios of hydrogen and lanthanum nuclei, respectively; $\tau$ is the decay time of the auto-correlation function and $E_a$ is the activation energy characterizing the mobility of hydrogen atoms. $M_{HH}$ and $M_{HM}$ are the second moments arising from dipolar interactions of pairs $H^1$ - $H^1$ and $H^1 - La^{139}$, respectively.

In (7) $\Sigma_j^M (R_{0j})^{-6}$ denotes the sum over the whole set of metal atoms surrounding the resonant H-atom located in the interstitial octa-position (0 0 0.5), or in the tetra-position (0.25 0.25 0.25). In (6) $\Sigma_j^{H(T)} (R_{0j})^{-6}$ denotes the sum over the totality of tetra-positions filled by $H_T$-atoms, while the sum $\Sigma_j^{H(O)} n_j (R_{0j})^{-6}$ includes the set of $H_O$-atoms. The factor $n_j$ describes the j-site occupation probability by one of $H_O$-atoms. In the case of a disordered state $n_j$ = const. = $c$, providing a linear dependence of $M_{HH}$ upon the $H_O$-atoms concentration $c$, but in the ordered phases $n_j$ is no more constant and becomes a function of the position number j, providing the dependence of $M_{HH}$ upon the hydrogen spatial distribution function $n(\mathbf{R}_j)$ and on the degree of order described by the temperature-dependent long-range-order parameters $\eta_1(T)$ and $\eta_2(T)$ [9]. The shape of the function $n(\mathbf{R}_j)$ in the case of $LaH_{2+c}$ compounds and the temperature dependence of corresponding equilibrium order parameters was defined in [9]. A second, indirect way of hydrogen ordering influence on the sum $\Sigma_j^{H(O)} n_j (R_{0j})^{-6}$ is associated with the distortion of the metal lattice caused by the redistribution of $H_O$-atoms, but the corresponding effect is significantly smaller than that induced by formation of a superstructure.

The sums $\Sigma_j^M (R_{0j})^{-6}$ and $\Sigma_j^{H(T)} (R_{0j})^{-6}$ are modified by the hydrogen ordering only indirectly, via distortions of the metal lattice. The lattice distortion effects will be neglected below.

Introducing in (6) and (7) the table-values of the constant parameters $\gamma_H$, $\gamma_M$ and h the following expressions are obtained:

$$M_{HH} = 358,4511 \ [\Sigma_j^{H(o)} n_j (R_{0j})^{-6} + \Sigma_j^{H(T)} (R_{0j})^{-6}] \ Oe \ , \quad (8)$$

$$M_{HM} = 66,7481 \ \Sigma_j^M (R_{0j})^{-6} \ Oe \ . \quad (9)$$

(Distances $R_{0j}$ are given in Å).

**Experimental results**
Measurements of spin-lattice relaxation times were carried out by spectrometer "Bruker SXP-100" on $f_0$ = 20 MHz frequency using $180° - \tau - 90°$ sequence of radio-frequency pulses. Main attention was paid to the low-temperature part of the $T_1(T)$ dependence in order to check the effects induced by hydrogen ordering and especially to look for the traces of the order-order type transformation. The hydrides were prepared and tested in E. Andronikashvili Institute of Physics.

In Fig.1 are presented temperature dependences of the hydrogen spin-lattice relaxation times $T_1$ (in msec units) measured on the samples $LaH_{2.34}$ and $LaH_{2.39}$. Temperatures are given in (1000 / T [K]) units. Along *y*-axes natural logarithms of relaxation times are measured.

**Results of analytical consideration**
*Location of the minimum of $T_{1d}$ (y) function.* Basing on formulae (1) – (7) an analytical expression of the first derivative $dT_{1d}(y)/dy$ can be obtained. On equaling it to zero and using the numerical values of parameters corresponding to $LaH_{2+c}$ system a following equation arise

$$R_{HM} [m_1(y) + 4 m_2(y)] + 0.1395 [m_3(y) + 3 m_1(y) + 6 m_4(y)] = 0, \quad (9)$$

where

$$R_{HM} = \Sigma_0^{HH} / \Sigma_0^{HM} \quad (10)$$

$$m_1(y) = (1-y)/(1+y^2)^2, \quad m_2(y) = (1-4y)/(1+4y^2)^2, \quad (11)$$

$$m_3(y) = (1-k_1 y)/(1+k_1 y^2)^2, \quad m_4(y) = (1-k_2 y)/(1+k_2 y^2)^2. \quad (12)$$

In (10) $\Sigma_0^{HH}$ and $\Sigma_0^{HM}$ denote sums $[\Sigma_j^{H(o)} n_j (R_{0j})^{-6} + \Sigma_j^{H(T)} (R_{0j})^{-6}]$ and $\Sigma_j^M (R_{0j})^{-6}$, respectively.

Equation (9) can be solved only numerically. Taking into account that the minimum of $T_{1d}$ (y) function is located within the disordered phases of both alloys (see Fig.A1) and introducing the values of parameters $R_{HM}$ ($LaH_{2.34}$) = 3.2233 and $R_{HM}$ ($LaH_{2.39}$) = 3.2370 (directly calculated for the given lattices) we obtain $y_{min}$($LaH_{2.34}$) = 0.6470 and $y_{min}$($LaH_{2.39}$) = 0.6468. Using these values and calculating $T_{1d}$ ($y_{min}$), on the basis of expressions (1) – (7) we find that

$$T_{1d} (LaH_{2.34})_{min} = 11.08 \text{ (msec)}, \quad T_{1d} (LaH_{2.39})_{min} = 10.99 \text{ (msec)}.$$

In both compounds measurements give $(T_{1d})_{min}$ = 20 (msec). Discrepancies between measured and calculated values of $(T_{1d})_{min}$ are of the order of magnitude registered already in [2] and in some other publications (see e.g. [11]).

*Temperature dependence of the spin-lattice relaxation time.* We have tried to describe the observed sequences of $T_{1d}$ (T) values given in the Fig.1 basing on expression (1) and on the values of second moments $M_{HH}$ and $M_{HV}$ calculated for the different states of hydrogen subsystem. It has to be noted that within the frames of the dipole-dipole interaction mechanism the spin-lattice relaxation time $T_{1d}$ is a function of the variable $y = \omega\tau$ and temperature dependence of $T_{1d}$ is defined by the temperature dependence of variable y, i.e. by expression (2).

It turns out to be impossible to describe the experimental curves in the whole temperature range with some constant values of activation energy $E_a$ and time parameter $\tau_0$. In these circumstances we have reanimated the idea of temperature dependent activation energy [2], adding the assumption of a temperature dependent time parameter $\tau_0$. Subdividing the temperature scale on a sequence of regions (see figs.2a-2b) and ascribing to each region the specific values of $E_a$ and $\tau_0$ (or $y_0$) parameters (see Table 1) we obtained the calculated $T_{1d}$ (T) dependences that practically coincide in the wide range of temperatures with the corresponding experimental curves (see figs. 3a-3b).

**Brief remarks**

**1.** Our consideration is based on the assumption that the mobility of hydrogen atoms can be described as an activation process characterized by corresponding energy and time constants (see exp. (2)). It must be noted as well that determined activation energies $E_a$ and time constants $\tau_0$ (or $y_0$) are of relative precision: the numerical values of these parameters depend on the selected positions of the temperature region limits (see Figs. 2a, 2b). Displacements of the region limits induce modifications of $E_a$ and $y_0$ values. As a result, a real physical sense can be ascribed only to the qualitative aspects of temperature dependence of these parameters.

**2.** Temperature induced modifications of $E_a$ and $y_0$ parameters are naturally associated with the ordering processes developed in the hydrogen subsystem. This idea is supported, first of all, by the changes of the $T_1(T)$ dependences in the vicinity of the theoretically defined phase transition points $T_{tr1}$ and $T_{tr2}$ in both samples (see Figs. 2 and Fig. A1). In the case of $LaH_{2.34}$ we have: $T_{tr1}$ = 398 K and $T_{tr2}$ = 330 K, and for the sample $LaH_{2.39}$ - $T_{tr1}$ = 402 K and $T_{tr2}$ = 235 K. (Additional information concerning the equilibrium ordered states at considered temperatures is given in Appendix). An analogous correlation between the phase transition temperatures and locations of the limits dividing the temperature scale into the regions associated with different activation energies was registered in $VH_x$ hydrides as well [12].

**3.** As the number of needed temperature region boundaries is more than phase transition numbers it follows that activation energy changes occur within the limits of the given ordered phases as well. From this detail it can be deduced that activation energy $E_a$ is apparently a smooth function of equilibrium long-range-order parameters $\eta_1(T)$ and $\eta_2(T)$ and that subdividing the temperature scale into the regions characterized by different $E_a$ values we try to reproduce this unknown $E_a(\eta_1(T), \eta_2(T))$ dependence.

## Conclusions

**1.** Flexions of the measured $T_1(T)$ dependences in the vicinity of the calculated temperatures Ttr$_2$ can be interpreted as a first, indirect confirmation of the existence of order-order transformations in LaH$_{2+c}$ hydrides, predicted in [9].

**2**. β-phases of rare-earth hydrides are not the best objects to demonstrate the influence of hydrogen ordering effects in the nuclear spin-lattice relaxation, as besides the contribution of the ordering subsystem of H$_O$-atoms a significant role plays the subsystem of H$_T$-atoms. Nevertheless, it is possible to register correlations between the shape changes of $T_1(T)$ dependences and the states of the ordering H$_O$-atoms.

**3**. The general tendency of activation energy reduction at lowering temperatures accompanied by the corresponding increase of time constant $\tau_0$, can be interpreted as an indication that at increase of the degree of order in the subsystem of H$_O$-atoms, collective modes of hydrogen motion become more significant than a diffusion based on uncorrelated jumps of individual particles.

**4.** In addition we have to note that activation energy decrease at temperature lowering demonstrated in the applied Table was already pointed out and briefly discussed in [13] on considering NMR in ScH$_x$, while the role of collective modes in the interstitial hydrogen motion was mentioned in [14].

**Acknowledgements**
The samples were prepared and tested by physicists of E.Andronikashvili Institute of Physics: N.Arabajian, A.Naskidashvili and V.Serdobintcev.

This research was supported by Award 3319 of Georgian Research and Development Foundation and U.S. Civil Research and Development Foundation.

**Appendix**
Experimental $T_1(T)$ dependencies presented in Fig.1 correspond to two lanthanum hydrides LaH$_x$ with "neighboring" hydrogen concentrations x = 2.34 and x = 2.39. Following previous experimental publications we could not expected essential differences neither in electronic structure, nor in metal lattice distortions of these alloys. Nevertheless, below 330 K (at [1000 / T] > 3) the corresponding relaxation time curves revealed some differences which we have ascribed to the influence of hydrogen ordering processes on the strength of dipole-dipole interaction. Rigorously, our analysis is not completely correct as we have not considered the alternative channels of the nuclear spin system energy dissipation (particularly, we did not estimated the role of conduction electrons and paramagnetic impurities). Below we give a brief comparative consideration of the role of the mentioned relaxation mechanisms.

The experimentally observed relaxation rate $R_1 = (1 / T_1)$ is usually presented as [4]

$$R_1 = R_{1p} + R_{1e} + R_{1d}, \qquad (A1)$$

where $R_{1p}$ describes the relaxation caused by interactions of nuclear-spin system with paramagnetic impurities, $R_{1e}$ - with conduction electrons of the hydride, and $R_{1d}$ - with the surrounding metal and hydrogen nuclei. It means that calculated relaxation time $T_1(T)$ presented in figures 3a and 3b had to be related not with the measured relaxation rate $R_1$, but with the difference $(R_1 - R_{1p} - R_{1e}) = R_{1d}$.

Let us consider the role of different relaxation channels separately.

*Influence of paramagnetic impurities.* It is expected usually that "proper lanthanum" samples contain an amount of paramagnetic cerium atoms, but both our hydrides were prepared from the same lanthanum ingot and their level of property was the same. Thus, it is difficult to conclude that the observed differences between $T_1(LaH_{2.34})$ and $T_1(LaH_{2.39})$ curves shown in Fig.1 can be related with this mechanism and we shall not consider it in details.

*Influence of conduction electrons.* The role of conduction electron subsystem is not so obvious. The corresponding relaxation rate $R_{1e}$ is presented usually as a ratio

$$R_{1e} = T / C_e, \qquad (A2)$$

where $C_e$ is the Korringa constant. Following [4] we have to assume that, in the case of $LaH_x$ system, $C_e$ is a function of hydrogen concentration x and can be approximated by the expression

$$C_e = A (3 - x)^{-(2/3)}. \qquad (A3)$$

The constant A can be deduced from $C_e(x)$ dependence (presented in the Fig.1 in [4]), and equals A = 330.2 [sec K], that for x = 2.27 provides the value $C_e$ = 407 [sec K], close to the value $C_e$ = 410 [sec K] used in [3] to describe very successfully the $T_1(T)$ dependence in "pure" $LaH_{2.27}$. Then basing on expression (A3) we obtain $C_e (LaH_{2.34})$ = 436 and $C_e (LaH_{2.39})$ = 460 [sec K].

In our analysis the relaxation time $T_1(T)$ we had ascribed entirely to the dipole-dipole interaction and within the frames of this assumption had determined the activation energy variations, subdividing the temperature scale on a number of intervals. Let us consider now the refinements that arise on taking into account the existence of the relaxation caused by the conduction electron subsystem.

Basing on the relation $R_{1d} = (R_1(exp) - R_{1e})$ and calculating the values of $R_{1e}$ at boundary temperatures T(bound) dividing the temperature regions, we can deduce the corrected values of relaxation times $T_{1d}$ denoted as $T_1$(exp, corrected) which are used for determination of the values of variable activation energy. Thus, we obtain:

in the case of $LaH_{2.34}$

| T(bound) [K] | 402 | 321 | 268 | 215 |
|---|---|---|---|---|
| $T_1$(exp) [msec] | 19.5 | 59 | 95 | 102 |
| $T_1$(exp, corrected) [msec] | 20 | 62 | 101 | 107 |

in the case of $LaH_{2.39}$

| T(bound) [K] | 435 | 343 | 292 | 243 |
|---|---|---|---|---|
| $T_1$(exp) [msec] | 20.5 | 58 | 81 | 95 |
| $T_1$(exp, corrected) [msec] | 21 | 61 | 85 | 100 |

It can be easily seen that in the determination of the activation energy values the corrections of measured relaxation times are of the minor significance and can not be considered to be responsible for the observed differences between experimental $T_1(LaH_{2.34})$ and $T_1(LaH_{2.39})$ curves.

*Influence of hydrogen ordering processes.* In spite of close hydrogen concentrations in $LaH_{2.34}$ and $LaH_{2.39}$ alloys the equilibrium ordered states in them are quite different in the temperature range under consideration. To illustrate this situation in Fig.A1 and Fig.A2 we had presented the phase diagram of La-H system (reproduced from I.G.Ratishvili, P.Vajda. Journ. All.Comp. 1997, v. 171, p. 252-254) and temperature dependences of equilibrium order parameters.

In the phase diagram are indicated the regions with different degree of ordering: "disordered states" (where both order parameters equal zero, $\eta_1 = 0$, $\eta_2 = 0$), partially ordered ("order I", $\eta_1 \neq 0$, $\eta_2 = 0$) and low-temperature ordered ("order II", $\eta_1 \neq 0$, $\eta_2 \neq 0$) states. It is significant that order-order transition line $T_{tr2}$ has a steep slope in the considered concentration range, that provides a significant difference between the $T_{tr2}$ ($LaH_{2.34}$) and $T_{tr2}$ ($LaH_{2.39}$) values. In this figure by the arrows are indicated concentrations under consideration and x = 2.27 investigated in [3]. A number of separate points represent the experimentally observed phase transitions (for details see [9]).

The differences between the ordered states formed in $LaH_{2.34}$ and $LaH_{2.39}$, respectively, are illustrated in Fig.A2, where are presented the temperature dependences of order parameters $\eta_1(T)$ and $\eta_2(T)$ for both alloys.

TABLE
**Activation energies $E_a$ and constants $y_0$ for the variable $y = \omega \tau$**

| LaH$_{2.34}$ |         | I (402-321) K | II (321-268) K | III (268-215) K |
|--------------|---------|---------------|----------------|-----------------|
|              | $E_a$ (eV) | 0.2686     | 0.0666         | 0.0072          |
|              | $y_0$   | 0.000277      | 0.41179        | 5.36189         |
| LaH$_{2.39}$ |         | I (435-343) K | II (343-292) K | III (292-243) K |
|              | $E_a$ (eV) | 0.2647     | 0.0565         | 0.0199          |
|              | $y_0$   | 0.000554      | 0.6350         | 2.7225          |

# FIGURE CAPTIONS

Fig. 1 The measured values of proton spin-lattice relaxation times in $LaH_{2.34}$ and $LaH_{2.39}$ hydrides.

Figs. 2 Subdivision of temperature scale into the regions of constant activation energies. The dashed flashes denote the boundaries of temperature regions. a – $LaH_{2.34}$, b – $LaH_{2.39}$.

Figs. 3 The measured and calculated values of proton spin-lattice relaxation times in lanthanum hydrides.

Open signs – calculated values, black signs – results of measurements. a – $LaH_{2.34}$, b – $LaH_{2.39}$.

Fig. A1  Phase diagram of $LaH_{2+c}$ system ($0.1 < c < 0.9$).

Fig. A2  Temperature dependences of equilibrium order parameters in lanthanum hydrides.

FIGURES

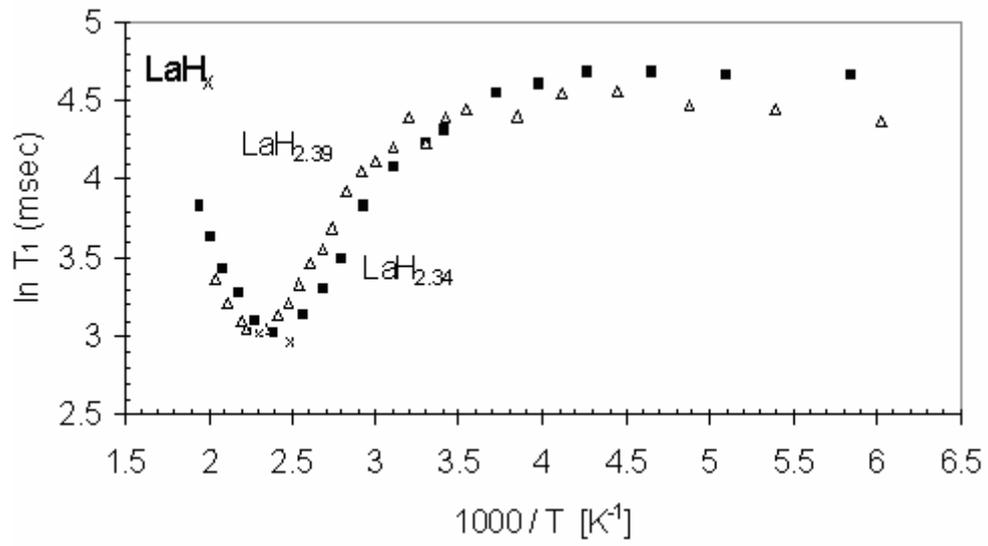

Fig.1.

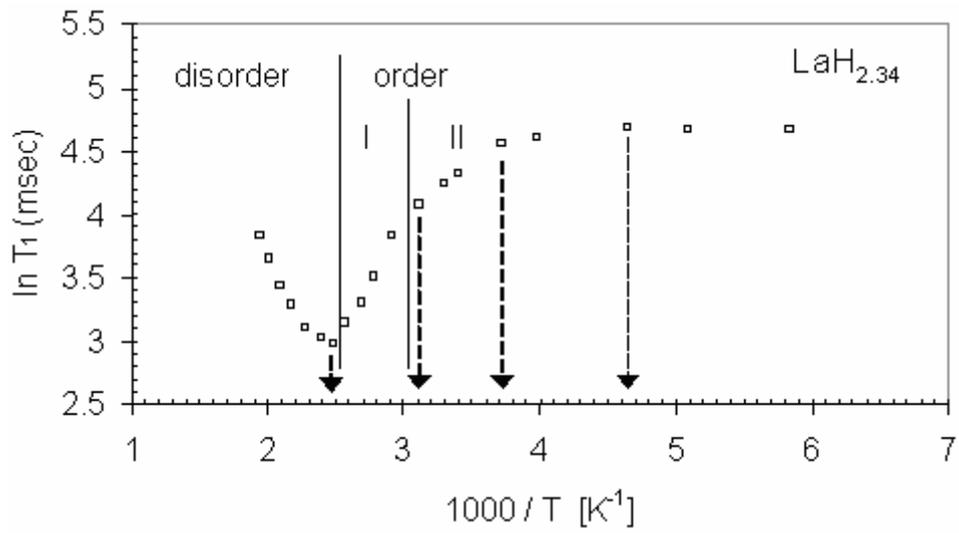

Fig. 2a

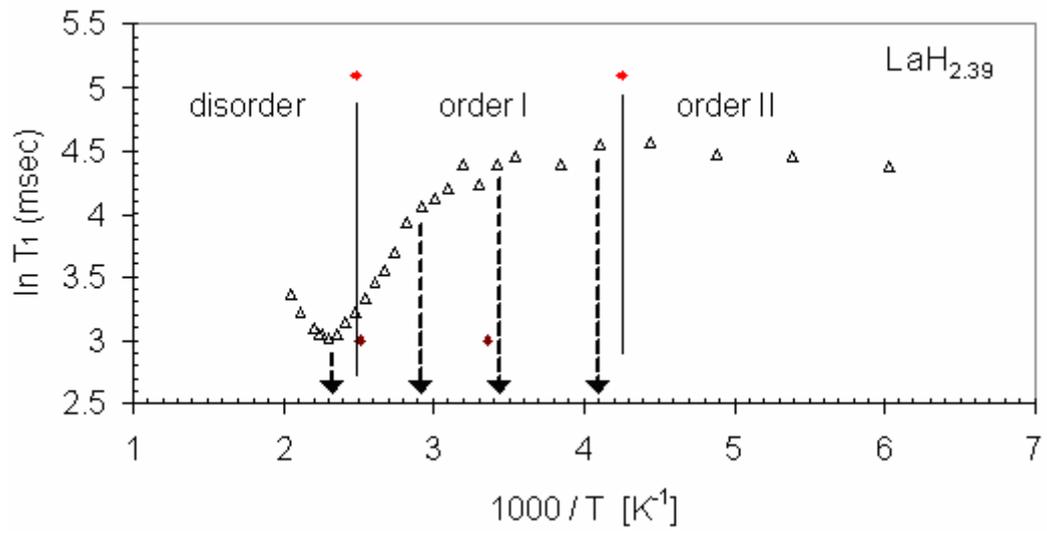

Fig. 2b

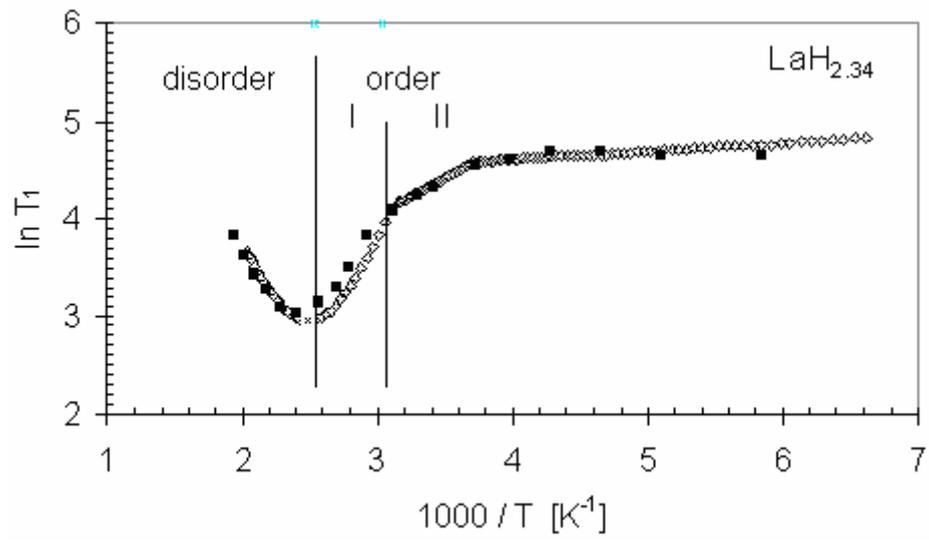

Fig. 3a

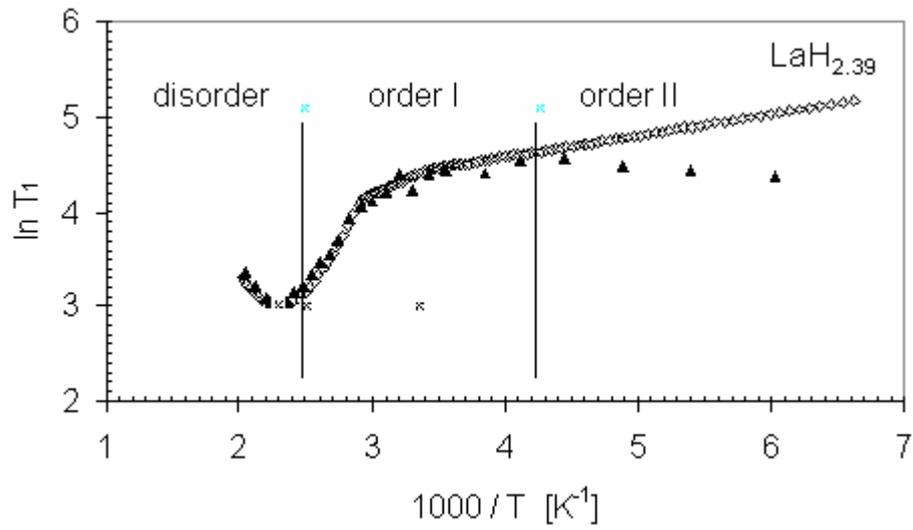

Fig. 3b

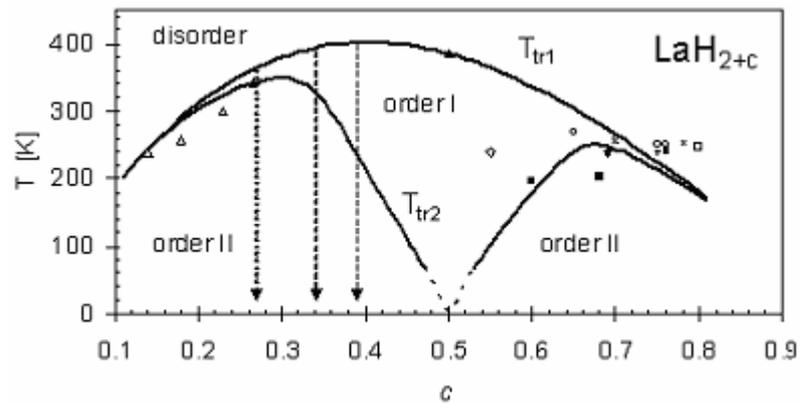

Fig. A1

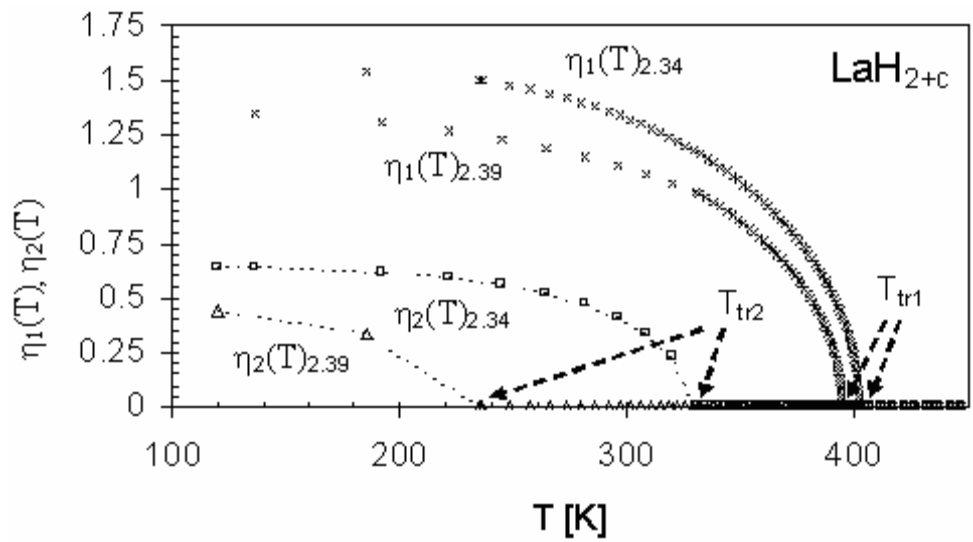

Fig. A2